\documentclass[prb,twocolumn,aps,showpacs,fixfloats]{revtex4}
\usepackage{graphicx}
\usepackage{bm}
\usepackage{amsmath,amssymb}
\usepackage{subfigure}
\usepackage{float}
\usepackage{latexsym}
\usepackage{color}
\usepackage{enumerate}
\usepackage{pdfpages}
\usepackage{tikz}
\usepackage{hyperref}
\DeclareMathSizes{10}{10}{7}{5}

\begin{document}
\newcommand{\s}{\scriptscriptstyle}
\newcommand{\uu}{\uparrow \uparrow}
\newcommand{\ud}{\uparrow \downarrow}
\newcommand{\du}{\downarrow \uparrow}
\newcommand{\dd}{\downarrow \downarrow}
\newcommand{\ket}[1] { \left|{#1}\right> }
\newcommand{\bra}[1] { \left<{#1}\right| }
\newcommand{\bracket}[2] {\left< \left. {#1} \right| {#2} \right>}
\newcommand{\vc}[1] {\ensuremath {\bm {#1}}}
\newcommand{\tr}{\text{Tr}}
\newcommand{\Trans}{\ensuremath \Upsilon}
\newcommand{\Refl}{\ensuremath \mathcal{R}}

\title{ Landau-Zener transition driven by a slow noise }

\author{Zhu-Xi Luo  and M. E. Raikh}

\affiliation{ Department of Physics and
Astronomy, University of Utah, Salt Lake City, UT 84112}

\begin{abstract}
The effect of a slow noise in non-diagonal matrix element, $J(t)$, that describes the diabatic level coupling, on the probability of the Landau-Zener transition is studied. For slow noise, the correlation time, $\tau_{\s c}$, of $J(t)$ is much longer than the characteristic time of the transition. Existing theory for this case suggests that the average transition probability is the result of averaging of the conventional Landau-Zener probability, calculated for a given constant $J$, over the distribution of $J$. We calculate a finite-$\tau_{\s c}$ correction to this classical result. Our main finding is that this correction is dominated by {\em sparse  realizations} of  noise for which $J(t)$ passes through zero within a narrow time interval near the level crossing. Two models of noise, random telegraph noise and gaussian noise, are considered. Naturally, in both models the average probability of transition decreases upon decreasing $\tau_{\s c}$. For gaussian noise we identify two domains of this fall-off with specific dependencies of average transition probability on $\tau_{\s c}$.
\end{abstract}

\maketitle

\section{Introduction}

A standard expression [\onlinecite{Landau1932,Zener1932,Majorana1932,Stueckelberg1933}] for the probability of the Landau-Zener transition between the two diabatic levels, $\pm vt/2$, reads
\begin{equation}
\label{PLZ}
P_{\s LZ}=1-Q_{\s LZ},  ~~~~Q_{\s LZ}= \exp\left(-\frac{2\pi J^2}{v}\right),
\end{equation}
where $J$ is the off-diagonal matrix element. Correspondingly, $Q_{\s LZ}$ is the probability to stay on the initial diabatic level.

The dynamics of the transition is governed by the following system
\begin{equation}
\label{system}
\begin{cases}
i{\dot a}_{\s 1}=\frac{vt}{2}a_1+Ja_2,  \\
i{\dot a}_{\s 2}=-\frac{vt}{2}a_2+Ja_1, \\
\end{cases}
\end{equation}
where $a_1$, $a_2$ denote the amplitudes for the particle to reside on the diabatic levels $1$ and $2$, respectively.

The question about the value of $P_{\s LZ}$ when the transition is driven by noise, so that $J$ is a random function of time,
was first put forward  and answered in Refs. [\onlinecite{Kayanuma1984,Kayanuma1985}].
The answer depends on whether the noise is slow or fast. More precisely, on whether the noise correlation time, $\tau_{\s c}$, is longer or shorter than the time of the Landau-Zener transition. For large matrix element, $J\gg v^{1/2}$, when $P_{\s LZ}$ is close to $1$, this transition time is given by
\begin{equation}
\label{tau1}
\tau_{\s LZ}=\frac{J}{v}.
\end{equation}

Fast noise corresponds to $\tau_{\s LZ}\gg \tau_{\s c}$, i.e. the matrix element oscillates many times in  the course of the transition. Then it is apparent that both outcomes of the transition are almost equally probable. In other words,  $Q_{\s LZ}$
 differs from  $1/2$ only slightly. It was demonstrated in Ref. [\onlinecite{Kayanuma1985}] that, for the fast noise,
\begin{equation}
\label{Q}
\langle Q_{\s LZ}\rangle =\frac{1}{2}\left[1+\exp{\left(-\frac{4\pi J_{\s c}^2}{v}\right)}\right],
\end{equation}
where $\langle ...\rangle$ denotes the averaging over the realizations of $J(t)$, and $J_{\s c}^2$ is defined via the noise correlator
\begin{equation}
\label{correlator}
\langle J(t)J(t')\rangle=J_{\s c}^2K(t-t')
\end{equation}
at coinciding times, where $K(0)=1$.

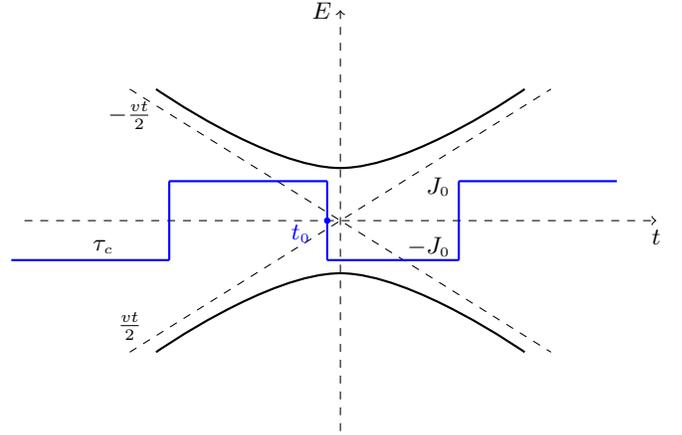
\begin{figure}[htbp]
	\centering
	\begin{tikzpicture}[scale=.7]
	\draw[dashed] (-4,2.5)--(4,-2.5);
	\draw[dashed] (-4,-2.5)--(4,2.5);
	\draw[dashed,->] (-6,0)--(6,0);
	\draw[dashed,->] (0,-4)--(0,4);
	
	\draw[thick]  plot[smooth, tension=.7] coordinates {(-3.5,2.5) (0,1) (3.5,2.5)};
	\draw[thick]  plot[smooth, tension=.7] coordinates {(-3.5,-2.5) (0,-1) (3.5,-2.5)};
	
	\draw[thick,blue] (-6.25,-0.75)--(-3.25,-0.75);
	\draw[thick,blue] (-3.25,-0.75)--(-3.25,.75);
	\draw[thick,blue] (-3.25,.75)--(-0.25,.75);
	\draw[thick,blue] (-0.25,.75)--(-0.25,-0.75);
	\draw[thick,blue](-0.25,-0.75)--(2.25,-0.75);
	\draw[thick,blue](2.25,-0.75)--(2.25,.75);
	\draw[thick,blue](2.25,.75)--(5.25,.75);
	
	\node at (-4,2) {$-\frac{vt}{2}$};
	\node at (-4,-2) {$\frac{vt}{2}$};
	\node at (-4.5,-0.5) {$\tau_{\s c}$};
	\node[below] at (6,0) {$t$};
	\filldraw[blue]  (-0.25,0) circle [radius=0.05];
	\node at (-0.75,-0.25) {\textcolor{blue}{$t_{\s 0}$}};
    \node[left] at (0,4) {$E$};
    \node[left] at (2.25,0.65) {$J_{\s 0}$};
    \node[left] at (2.25,-0.5) {$-J_{\s 0}$};
	\end{tikzpicture}
	\caption{Schematic illustration of the Landau-Zener transition driven by slow telegraph noise. For a typical noise realization
the matrix element switches from $J_{\s 0}$ to $-J_{\s 0}$ at time $t_{\s 0} \sim \tau_{\s c} \gg \tau_{\s LZ}$. For such realizations
the probability, $Q_{\s LZ}$, to stay on the same diabatic level is exponentially small. However, for sparse realizations with
$t_{\s 0} \lesssim \tau_{\s LZ}$, the value $Q_{\s LZ}$ is close to $1$. Thus, it is these sparse realizations are responsible for the finite-$\tau_{\s c}$ correction.   }
	\label{Fig1}
\end{figure}

In the opposite limit of slow noise $\tau_{\s c} \gg \tau_{\s LZ}$, the result obtained in Ref. [\onlinecite{Kayanuma1985}] is also physically transparent. Namely, one can neglect the change of $J(t)$
during the transition. This suggests that the expression Eq. \eqref{PLZ}
should be simply averaged over the distribution, ${\cal P}(J)$, of the matrix element
\begin{equation}
\label{averaged}
\langle Q_{\s LZ}\rangle =\int\limits_{-\infty}^{\infty}dJ~ {\cal P}(J) \exp{\left\{-\frac{2\pi J^2}{v}\right\}}.
\end{equation}

Further developments in the theory of the noise-driven Landau-Zener transition in Refs.  [\onlinecite{Kayanuma1987,Kayanuma1998,Kayanuma2007,Ao1989,Ao1991,Pokrovsky2003,Sinitsyn2003,Pokrovsky2004-1,Pokrovsky2004-2,Pokrovsky2007,Galperin2008,Imambekov2010, Imambekov2013, Ziman2011,Fai2013,Fai2015,Fai2016}] included: (\romannumeral 1) consideration of specific microscopic models of the environment leading to random $J(t)$; (\romannumeral 2) extension of Eq. (\ref{Q}) to the case when $J(t)$ has both constant and fluctuating components (Pokrovsky-Sinitsyn formula); and (\romannumeral 3) generalization of the theory of the fast noise to the case of three and more crossing levels. 

With regard to the papers [\onlinecite{Kayanuma1987,Kayanuma1998,Kayanuma2007,Ao1989,Ao1991,Pokrovsky2003,Sinitsyn2003,Pokrovsky2004-1,Pokrovsky2004-2,Pokrovsky2007,Galperin2008,Imambekov2010, Imambekov2013, Ziman2011,Fai2013,Fai2015,Fai2016}], we note that the results for the average transition probabilities were obtained only in two ``extreme" cases:
$\tau_{\s c} \rightarrow 0$ for the fast noise and $\tau_{\s c} \rightarrow \infty$ for the slow noise. The corrections in small parameters
$\tau_{\s c}/\tau_{\s LZ}$ for the fast noise and $\tau_{\s LZ}/\tau_{\s c}$ for the slow noise were not found.
There is actually a fundamental reason for this, which can be traced back to the techniques employed in Ref. [\onlinecite{Kayanuma1985}] and subsequent works.
Namely, in  Ref. [\onlinecite{Kayanuma1985}], $Q_{\s LZ}$ was presented as expansion in powers of $J^2$, and then averaged over random
realizations of $J(t)$ term-by-term. As we will see below, this technique does not allow to describe a crossover between the limits Eqs. (\ref{Q})  and (\ref{averaged}) and even to capture finite-$\tau_{\s c}$ corrections.

Calculation of a finite-$\tau_{\s c}$ correction for the case of a slow noise is the main subject of the present paper. We will demonstrate
that this correction comes from {\em non-perturbative} effects, or in other words, not from typical, but rather from  sparse noise realizations. To illustrate the message,
consider the particular model of telegraph noise as in Ref. [\onlinecite{Galperin2008}], when $J(t)$ switches randomly between the two values $\pm J_{\s 0}$. We will demonstrate that it is particular the
switchings which take place near the level crossing  $t=0$, that have a dramatic effect on $Q_{\s LZ}$.  This situation is illustrated in Fig.
\ref{Fig1}.

Without switching near $t=0$, the probability $Q_{\s LZ}$ is exponentially small. On the other hand, {\em with switching},
as we will demonstrate below, $Q_{\s LZ}$ is close to $1$. Since the probability that the switching takes place at $t_{\s 0} \lesssim \tau_{\s LZ}$ is $\sim \tau_{\s LZ}/\tau_{\s c}$, the correction to $Q_{\s LZ}$ can be estimated as $\sim \tau_{\s LZ}/\tau_{\s c}$. This correction becomes important even at large enough $\tau_{\s c}$, since in the limit $\tau_{\s c} \rightarrow \infty$ the value $Q_{\s LZ}$ is exponentially small.
In the next Section we justify the above  picture by a rigorous calculation. In Sect. \ref{Sec:Gaussian} we study the case of a gaussian noise when $J(t)$ is a smooth function of time. We show that in this case the finite-$\tau_{\s c}$ correction to $Q_{\s LZ}$, again, originates from the sparse realizations of noise when $J(t)$ passes through zero near the moment of the level crossing. In addition to finite-$\tau_{\s c}$ correction, the realizations with $J(t_{\s 0})=0$ for $t_{\s 0} \ll \tau_{\s c}$ determine  the behavior of  $Q_{\s LZ}$ upon decreasing of the correlation time. We identify two domains of $\tau_{\s c}$ with distinctively different $Q_{\s LZ}(\tau_{\s c})$~-~dependence. In Sect. \ref{Sec:Summary}, which concludes the paper, we speculate on the form of the finite-$\tau_{\s c}$ correction to $Q_{\s LZ}$ in the limit of fast noise $\tau_{\s c} \ll \tau_{\s LZ}$.

\section{Telegraph noise: switching near $t=0$}\label{Sec:Telegraph}

Consider the system Eq. \eqref{system} and assume that the coupling constant is equal to $J_{\s 0}$ for $t<t_{\s 0}$, while for $t>t_{\s 0}$ it switches to $-J_{\s 0}$, see Fig. \ref{Fig1}. Solutions of the system for $t<t_{\s 0}$ are expressed via the parabolic cylinder functions [\onlinecite{Bateman}] as follows
\begin{equation}
\label{solution1}
\begin{cases}
a_1=D_\nu(z),\\
a_2=-i\sqrt{\nu} D_{\nu-1}(z),
\end{cases}
\end{equation}
where the argument $z$ is defined as $z=\sqrt{v}e^{i\pi/4}t$ and
the index $\nu$ is given by
\begin{equation}
\label{nu}
\nu=-\frac{iJ^2}{v}.
\end{equation}
The solution Eq. \eqref{solution1} ensures that $a_2(-\infty)=0$,
i.e. that the particle is in the state $a_1$ away from the crossing.

For $t>t_{\s 0}$ solution of the system represents a linear combination of
two parabolic-cylinder functions
\begin{equation}
\label{solution2}
\begin{cases}
a_1=A D_\nu(z) +B D_\nu (-z),\\
a_2=i\sqrt{\nu} A D_{\nu-1}(z)-i\sqrt{\nu} B D_{\nu-1} (-z).\\
\end{cases}
\end{equation}
The fact that the coefficient in front of $D_{\nu-1}(z)$ in $a_2$
is equal to $-A$, unlike Eq. \eqref{solution1}, is the consequence of the switching. The coefficients
$A$ and $B$ are found from continuity of $a_1$ and $a_2$ at $t=t_{\s 0}$.
The corresponding system reads
\begin{equation}
\label{eq:TeleSystem}
\begin{cases}
D_\nu(z_{\s 0})=A D_\nu(z_{\s 0}) +B D_\nu (-z_{\s 0}), \\
D_{\nu-1}(z_{\s 0})=-A D_{\nu-1}(z_{\s 0}) +B D_{\nu-1} (-z_{\s 0}),\\
\end{cases}
\end{equation}
where $z_{\s 0}=\sqrt{v}e^{i\pi/4}t_{\s 0}$. From this system we readily derive the following expressions
\begin{equation}
\label{eq:AB}
A=\frac{\frac{D_{\nu-1}(-z_{\s 0})}{D_{\nu-1}(z_{\s 0})}-\frac{D_\nu(-z_{\s 0})}{D_\nu(z_{\s 0})}}{\frac{D_{\nu-1}(-z_{\s 0})}{D_{\nu-1}(z_{\s 0})}+\frac{D_\nu(-z_{\s 0})}{D_\nu(z_{\s 0})}},
~~B=\frac{2}{\frac{D_{\nu-1}(-z_{\s 0})}{D_{\nu-1}(z_{\s 0})}+\frac{D_\nu(-z_{\s 0})}{D_\nu(z_{\s 0})}}.
\end{equation}
Without switching, we would have $A=1$, $B=0$. Suppose now, that the switching took place at $t_{\s 0}=0$. Then from
Eq. \eqref{eq:AB} we have $A=0$ and $B=1$. This illustrates the dramatic effect of switching on the probability $Q_{\s LZ}$. Indeed, in the limit $z_{\s 0}\rightarrow \infty$, the ratio $|D_{\nu}(z_{\s 0})/D_{\nu}(-z_{\s 0})|^2$ approaches a small value $\exp(-2\pi J_{\s 0}^2/v)$, which is the value of $Q_{\s LZ}$ without switching. By contrast, with $B=1$ as a result of switching, $Q_{\s LZ}$ is equal to $1$.

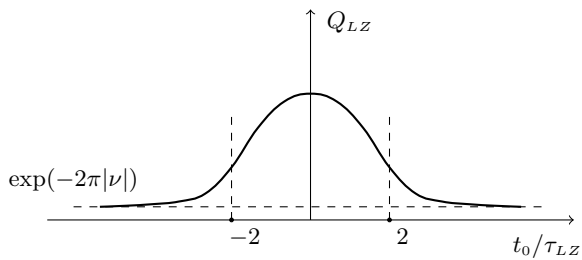
\begin{figure}[htbp]
\centering
\begin{tikzpicture}[scale=.7]
\draw[->] (-5,0)--(5,0);
\draw[->] (0,0)--(0,4);
\draw[thick] plot[smooth, tension=.7] coordinates {(-4,0.25) (-3.5,0.27) (-3,0.3)
(-2.5,0.35) (-2,0.5) (-1.5,1) (-1,1.75) (-0.5,2.25) (0,2.4) (0.5,2.25) (1,1.75) (1.5,1)
(2,0.5) (2.5,0.35) (3,0.3) (3.5,0.27) (4,0.25)};
\draw[dashed] (-1.5,0)--(-1.5,2);
\draw[dashed] (1.5,0)--(1.5,2);
\draw[dashed] (-4.5,0.25)--(4.5,0.25);
\node at (0.75,3.75) {$Q_{\s LZ}$};
\node at (-4.5,0.75) {$\exp(-2\pi|\nu|)$};
\node at (4.5,-0.5) {$t_{\s 0}/\tau_{\s LZ}$};
\node [below] at (-1.25,0) {$-2$};
\node [below] at (1.75,0) {$2$};
\filldraw (-1.5,0) circle (1pt) ;
\filldraw (1.5,0) circle (1pt);
\end{tikzpicture}
\caption{The dependence of the probability to stay on the initial diabatic level at $t\rightarrow \infty$ is
 shown schematically versus the moment of switching, $t_{\s 0}$.
The probability $Q_{\s LZ}(t_{\s 0})$ is described by the Lorentizian Eq. (\ref{eq:Qtilde-1}) for $t_{\s 0}\sim \tau_{\s LZ}$ and
approaches asymptotically to $\exp(-2\pi|\nu|)$ for $t_{\s 0}$ much bigger than the time, $\tau_{\s LZ}$, of the Landau-Zener transition. }
\label{figN}
\end{figure}

The situation $t_{\s 0}=0$ is most favorable for enhancement of $Q_{\s LZ}$. In order to find the average enhancement, we should analyze the
expressions for $A$ and $B$, given by Eq. \eqref{eq:AB}, as a function of $z_{\s 0}$.
The fact which allows to carry out this analysis analytically is that the typical value of $Q_{\s LZ}$ is small. This is also equivalent to the condition $J_{\s 0}^2 \gg v$, or $|\nu|\gg 1$. Conventionally, the behavior of the parabolic cylinder functions at large $|z_{\s 0}| \gg 1$ is obtained from the semiclassics. The condition $|\nu|\gg 1$ justifies the semiclassical analysis even for small $z_{\s 0}$. This analysis is carried out in Appendix \ref{Sec:App}. Substituting Eqs.\eqref{8} and \eqref{10} into Eq. \eqref{eq:AB}, we obtain the following expression for $B$

\begin{widetext}
\begin{equation}
\label{eq:B}
B=2\left[\frac{\left(\frac{\sqrt{4\nu-z_{\s 0}^2}-iz_0}{\sqrt{4\nu-z_{\s 0}^2}+iz_{\s 0}}\right)^{\nu+1/2}+e^{i\nu\pi+\frac{iz_{\s 0}}{2}\sqrt{4\nu-z_{\s 0}^2}}}{e^{\frac{iz_{\s 0}}{2}\sqrt{4\nu-z_{\s 0}^2}}+e^{i\nu\pi}\left(\frac{\sqrt{4\nu-z_{\s 0}^2}-iz_{\s 0}}{\sqrt{4\nu-z_{\s 0}^2}+iz_{\s 0}}\right)^{\nu+1/2}}+\frac{\left(\frac{\sqrt{4\nu-z_{\s 0}^2}-iz_{\s 0}}{\sqrt{4\nu-z_{\s 0}^2}+iz_{\s 0}}\right)^{\nu-1/2}-e^{i\nu\pi+\frac{iz_{\s 0}}{2}\sqrt{4\nu-z_{\s 0}^2}}}{e^{\frac{iz_{\s 0}}{2}\sqrt{4\nu-z_{\s 0}^2}}-e^{i\nu\pi}\left(\frac{\sqrt{4\nu-z_{\s 0}^2}-iz_{\s 0}}{\sqrt{4\nu-z_{\s 0}^2}+iz_{\s 0}}\right)^{\nu-1/2}}\right]^{-1}.
\end{equation}
\end{widetext}

The above expression for $B$ allows a dramatic simplification. It is achieved upon adding the two fractions and subsequently rewriting the result in the form of a single fraction. The result reads
\begin{equation}
\label{eq:SimpB}
B=\frac{iz_{\s 0}+i\sqrt{4\nu}\sin(\phi+\nu\pi)}{\sqrt{4\nu-z_{\s 0}^2}\sinh(|\nu|\pi)},
\end{equation}
where the phase $\phi$ is defined as
\begin{equation}
\label{eq:phi}
\phi=-\frac{z_{\s 0}}{2}\sqrt{4\nu-z_{\s 0}^2}-i\nu\ln \frac{(\sqrt{4\nu-z_{\s 0}^2}-iz_{\s 0})^2}{4\nu}.
\end{equation}

In a similar way, from Eq. \eqref{eq:AB} we get a simplified expression for $A$ 
\begin{equation}
\label{eq:SimpA}
A=-\frac{i\sqrt{4\nu}\sin\phi+iz_{\s 0}\cosh(|\nu|\pi)}{\sqrt{4\nu-z_{\s 0}^2}\sinh(|\nu|\pi)}.
\end{equation}

The exact expression for the probability to stay on the same diabatic level in the presence of
the switching is ${Q}_{\s LZ}=|a_1(\infty)/a_1(-\infty)|^2$. From Eq. (\ref{solution2})
we can express this probability in terms of $A$ and $B$ as follows
\begin{equation}
\label{eq:Qtilde}
{Q}_{\s LZ}=|A|^2 e^{-2\pi|\nu|}+(A^*B+B^*A)e^{-\pi|\nu|}+|B|^2.
\end{equation}
Let us first check that for large $|z_{\s 0}|\gg |\nu|^{1/2}$ Eq. \eqref{eq:Qtilde} reproduces ${Q}_{\s LZ}=\exp(-2\pi|\nu|)$, as is expected on physical grounds. Indeed, in this limit, we can replace $A$ by $-1$ and $B$ by $2\exp(-\pi|\nu|)$. Then the second and the third terms in Eq. \eqref{eq:Qtilde} cancel out.

To study the behavior of ${Q}_{\s LZ}$ in the domain of $|z_{\s 0}|\sim |\nu|^{1/2}$, we notice that only the third term does not contain any exponentially small factor. Hence, for this term, we can use the asymptotic expression in which corrections of the order $\exp(-\pi|\nu|)$ are neglected. Namely, 
$B\approx \left(z_{\s 0}^2/4\nu-1\right)^{-1/2} e^{-i\phi}$. This immediately leads to the following expression for ${Q}_{\s LZ}$, in which we return to the original notations $t_{\s 0}$ and $\tau_{\s LZ}$,
\begin{equation}
\label{eq:Qtilde-1}
{Q}_{\s LZ}=\frac{4}{4+\left(\frac{t_{\s 0}}{\tau_{\s LZ}}\right)^2}.
\end{equation}
In Fig. \ref{figN} the dependence of ${Q}_{\s LZ}$ on the moment of switching is shown schematically. It is described by the Lorentizian Eq. (\ref{eq:Qtilde-1}) for $t_{\s 0}\sim \tau_{\s LZ}$ and
approaches $\exp(-2\pi|\nu|)$ for $t_{\s 0}\gg\tau_{\s LZ}$.

As we established above, the switching at $t_{\s 0}=0$ leads to
the probability ${Q}_{\s LZ}=1$. Then Eq. \eqref{eq:Qtilde-1}
suggests that the enhancement of ${Q}_{\s LZ}$ takes place
in the domain of $t_{\s 0}\sim \pm 2\tau_{\s LZ}$. Since $t_{\s 0}$
is random, we average Eq. \eqref{eq:Qtilde-1} over $t_{\s 0}$ to
get the net enhancement of ${Q}_{\s LZ}$,
and present the final result in the form
\begin{equation}
\label{eq:TeleAverage}
\begin{split}
1-P_{\s LZ}& = \exp\left\{-\frac{2\pi J^2}{v}\right\}+\frac{1}{\tau_{\s c}}\int_{-\infty}^{\infty}dt_{\s 0}\ \frac{4}{4+\left(\frac{t_{\s 0}}{\tau_{\s LZ}}\right)^2}\\
& = \exp\left\{-\frac{2\pi J^2}{v}\right\}+2\pi\frac{\tau_{\s LZ}}{\tau_{\s c}}. \\
\end{split}
\end{equation}
It is important to note that the condition of applicability of Eq. \eqref{eq:Qtilde-1} is $\tau_{\s LZ}\ll \tau_{\s c}$. Under this condition, the second term of Eq. \eqref{eq:TeleAverage} can greatly exceed the first term, which is the result corresponding to no switching.

\section{Gaussian noise}\label{Sec:Gaussian}
\begin{figure}[htbp]
\centering
\begin{tikzpicture}[scale=.7]
\draw[dashed] (-4,2.5)--(4,-2.5);
\draw[dashed] (-4,-2.5)--(4,2.5);
\draw[->] (-6,0)--(6,0);
\draw[->] (0,-4)--(0,4);
\draw[thick]  plot[smooth, tension=.7] coordinates {(-3.5,3.0) (0.5,0.8) (4,3.5)};
\draw[thick]  plot[smooth, tension=.7] coordinates {(-3.5,-3) (0.5,-0.8)(4,-3.5)};
\draw[blue,thick] (-5,0) sin (-2,.5) cos (1,0) sin (4,-0.5);
\draw[dotted, thick] (1,-3)--(1,3);
\node at (-4.5,2) {$-\frac{vt}{2}$};
\node at (-4.5,-2) {$\frac{vt}{2}$};
\node at (-4.5,-0.5) {$\tau_{\s c}$};
\node[below] at (6,0) {$t$};
\filldraw[thick,blue]  (1,0) circle [radius=0.05];
\node[right] at (1,-0.25) {$t_{\s 0}$};
\end{tikzpicture}
\caption{(Color online). If the matrix element, $J(t)$, turns to zero at a some $t=t_{\s 0}$ close to the crossing of the
diabatic levels (blue curve),  the calculation of the transition probability reduces to the Landau-Zener problem with renormalized velocity and
coupling both depending on $t_{\s 0}$. The moment $t_{\s 0}$ also  defines the shift of minima of adiabatic levels from $t=0$. This
 shift is smaller that $t_{\s 0}$, see Eq. (\ref{b-eq-phi}). }
\label{continuous}
\end{figure}
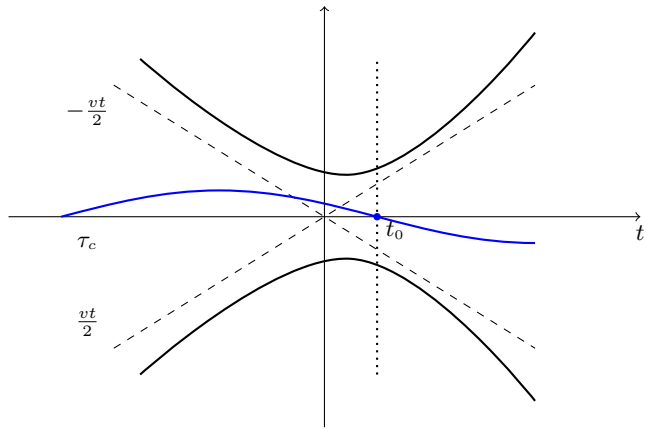

Suppose that $J(t)$ changes continuously as in Fig. \ref{continuous} and that the typical $J$ is much bigger than $v^{1/2}$, so that the typical $P_{\s LZ}$ is close to $1$. As $J$ slowly changes with time, the maximum contribution to $Q_{\s LZ}$ will come from the time domain when it takes small values
$J\lesssim v^{1/2}$. Since the portion of these domains is $\sim v^{1/2}/J$, the ratio $v^{1/2}/J$   determines the average $Q_{\s LZ}$. This is how the analytical result of Ref. [\onlinecite{Kayanuma1985}]
can be interpreted. This result applies for very long correlation times, $\tau_{\s c}\rightarrow \infty$.

As we are interested in a finite-$\tau_{\s c}$ correction, we will study the domains of small $J$
in more detail. A nontrivial consequence of small $J$ values is that, when the noise takes these
values, the Landau-Zener transition time is much shorter than the typical $\tau_{\s LZ}$. This, in turn,
suggests that even when the condition of slow noise is violated for typical $J$, it can still
be met for anomalously small $J$, which determines $Q_{\s LZ}$. Thus, one can expect a
nontrivial behavior of $Q_{\s LZ}$ upon decreasing the correlation time.

Naturally, small $J$-values are realized in the vicinity of zeros of $J(t)$. Similar to
the previous Section, the most relevant are the realizations of noise when a zero occurs in
the vicinity of $t=0$, when the levels cross. For these realizations we can linearize $J(t)$
as follows
\begin{equation}
\label{Prime}
J(t)=(t-t_{\s 0})J',
\end{equation}
where $J'$ is the slope and $t_{\s 0}$ is much smaller than the correlation time.

Our prime observation is that the system Eq. \eqref{system} can be solved exactly with $J(t)$ in the form Eq. (\ref{Prime}). This is achieved by introducing new variables
\begin{equation}
\label{rotation}
\begin{cases}
b_1=a_1\cos\varphi+a_2\sin\varphi,\\
b_2=a_1\sin\varphi-a_2\cos\varphi,\\
\end{cases}
\end{equation}
where the angle $\varphi$ is defined as
\begin{equation}
\label{phi}
\tan (2\varphi) = \frac{2J'}{v}.
\end{equation}
It is straightforward to check that the system of equations for $b_1$, $b_2$ has the form
\begin{equation}
\label{b-eq-phi}
\begin{cases}
i\dot{b}_{\s 1}=\frac{v}{2} \frac{(t-t_{\s 0}\sin^2(2\varphi))}{\cos (2\varphi)} \ b_1+ \frac{v}{2} t_{\s 0} \sin (2\varphi)\ b_2,\\
i\dot{b}_{\s 2}=-\frac{v}{2} \frac{(t-t_{\s 0}\sin^2(2\varphi))}{\cos (2\varphi)} \ b_2+ \frac{v}{2} t_{\s 0} \sin (2\varphi)\ b_1.\\
\end{cases}
\end{equation}
We see that the system has reduced to the original system Eq. \eqref{system}
with renormalized velocity ${\tilde v}=v/\cos(2\varphi)$ and the coupling ${\tilde J}=vt_{\s 0}\sin(2\varphi)/2$.
The established mapping allows us to write the answer for $Q_{\s LZ}$ straightaway[\onlinecite{footnote}]
\begin{equation}
\label{nothing}
Q_{\s LZ}(t_{\s 0})=\exp\left\{-\frac{2\pi J'^2t_{\s 0}^2}{v}\left|\cos^3 (2\varphi)\right|\right\}.
\end{equation}
Using Eq. \eqref{phi}, the above expression can be recast into the form
\begin{equation}
\label{Q-LZ}
Q_{\s LZ}=\exp\left\{-\frac{2\pi J'^2t_{\s 0}^2}{v}\left(\frac{v^2}{v^2+4J'^2}\right)^{3/2}\right\}.
\end{equation}
The crucial assumption made in course of deriving Eq. (\ref{Q-LZ}) is that the linearization Eq. (\ref{Prime}) is valid during the entire  renormalized
transition time, ${\tilde \tau}_{\s LZ}$. We are now in position to check this assumption. Indeed,
\begin{equation}
{\tilde \tau}_{\s LZ}=\frac{\tilde{J}}{\tilde{v}}=\frac{J'v}{v^2+4J'^2}t_{\s 0}.
\end{equation}
Since the first factor does not exceed $1$, we conclude that the condition $t_{\s 0} \ll \tau_{\s c}$ is the only condition
necessary for Landau-Zener transition to be dominated by a local zero of $J(t)$.

Contribution of the realizations with $t_{\s 0}\ll \tau_{\s LZ}$ to the
probability $Q_{\s LZ}$ is given by the average
\begin{equation}
\label{Average}
\overline{ Q}_{\s LZ} =\int_{-\infty}^{\infty}\frac{dt_{\s 0}}{\tau_{\s c}} Q_{\s LZ}(t_{\s 0})=\left(\frac{v}{2}\right)^{1/2}\frac{1}{|J'|\tau_{\s c}}\left(\frac{v^2+4J'^2}{v^2}\right)^{3/4}.
\end{equation}
We are interested in $\tau_{\s c}$-dependence of the probability $Q_{\s LZ}$.
The correlation time enters into Eq. (\ref{Average}) in two ways: firstly it is present directly
in the denominator, and, secondly, via $J'$, since its typical value is $J'\sim J_{\s c}/\tau_{\s c}$.
The remaining task is to average Eq. (\ref{Average}) over the random
slopes, $J'$. We will perform this averaging in the domains of small and large $J'$ separately:

(\romannumeral 1) Small $J'$. In this limit, we can replace the bracket in Eq. (\ref{Average}) by $1$. The average over $J'$ does not diverge, since $J'$ cannot be smaller than the minimal value determined by the condition $t_{\s 0} \ll \tau_{\s c}$. Since the typical $t_{\s 0}$ is $\sim v^{1/2}/J'$, the estimate for the minimal $J'$ is $J'<v^{1/2}/\tau_{\s c}$. This minimal $J'$ fixes the lower
limit in the integral
\begin{equation}
\label{AverageAverage}
\langle \overline{Q}_{\s LZ}\rangle= \left(\frac{v}{\pi}\right)^{1/2}\frac{1}{J_{\s c}'\tau_{\s c}}\int_{v^{1/2}/\tau_{\s c}}^{\infty}\frac{dJ'}{J'}
\exp{\Big(-\frac{J'^2}{2J_{\s c}'^2}}\Big),
\end{equation}
where $J_{\s c}'$ is the width of the Gaussian distribution of $J'$,
\begin{equation}
\label{Jc}
J_{\s c}'^2=J_{\s c}^2\frac{\partial^2 K}{\partial t_{\s 1}^2}\Big|_{t_{\s 1}=t_{\s 2}},
\end{equation}
 and the correlator $K$ is defined in Eq. \eqref{correlator}. Within a number under the logarithm, the final result for the double average reads
\begin{equation}
\label{Average-J}
\langle \overline{Q}_{\s LZ}\rangle =\left(\frac{v}{\pi}\right)^{1/2}\frac{1}{J_{\s c}'\tau_{\s c}}
\ln\Big(\frac{J_{\s c}'\tau_{\s c}}{v^{1/2}}\Big).
\end{equation}
Let us compare this result with a standard expression derived in Ref. [\onlinecite{Kayanuma1985}].
In our limit, $J_{\s c}\gg v^{1/2}$, the result of gaussian averaging in Eq. (\ref{averaged}) reads
\begin{equation}
\label{1985}
\langle Q_{\s LZ}\rangle =\left(\frac{v}{4\pi}\right)^{1/2}\frac{1}{J_{\s c}}.
\end{equation}
Since $J_{\s c}' \sim J_{\s c}/\tau_{\s c}$, the prefactor in Eq. (\ref{Average-J}) is of the same order as in Eq. (\ref{1985}).
In particular case of Gaussian correlator, we have $J_{\s c}'\tau{\s c}=2^{1/2}J_{\s c}$, so that the prefactor in Eq. (\ref{Average-J}) is two times bigger than in Eq. (\ref{1985}). However, our result contains an additional big factor $\ln(J_{\s c}/v^{1/2})$. This factor originates from realizations of noise for which $J(t)$ passes through zero near $t=0$.
To the best of our knowledge, the importance of these realizations leading to the enhancement of $Q_{\s LZ}$  was not mentioned in the literature.

The result Eq. (\ref{Average-J}) yields the average $Q_{\s LZ}$ in the
limit $\tau_{\s c}\rightarrow \infty$. Calculation of the finite-$\tau_{\s c}$ correction is
straightforward. Expanding the bracket in Eq. (\ref{Average})
to the first power in the small parameter $J'^2/v^2$, and averaging over $J'$, we get
\begin{equation}
\label{Qcorrection}
 \langle\overline{Q}_{\s LZ}(\tau_{\s c})\rangle-\langle\overline{Q}_{\s LZ}(\infty)\rangle
\!=\! \frac{3}{(\pi v^3)^{1/2}}
\!\int\limits_{0}^{\infty}\! dJ'\!~
\frac{|J'|}{\tau_{\s c}J_{\s c}'}\exp\left(-\frac{J'^2}{2J_{\s c}'^2}\right).
\end{equation}
It follows from Eq. (\ref{Qcorrection}) that upon decreasing the correlation
time the correction to $\overline{Q}_{\s LZ}$
grows as $1/\tau_{\s c}^2$. It is instructive
to rewrite this correction as
\begin{equation}
\label{Qcorrection1}
\langle\overline{Q}_{\s LZ}(\tau_{\s c})\rangle-\langle\overline{Q}_{\s LZ}(\infty)\rangle
= 3 \left(\frac{2}{\pi}\right)^{1/2}\left(\frac{J_{\s c}}{v^{1/2}}\right)\left(\frac{1}{v^{1/2}\tau_{\s c}}\right)^2.
\end{equation}
Naturally, the applicability of Eq. (\ref{Qcorrection1}) is
terminated as $\tau_{\s c}$ becomes smaller than $J_{\s c}/v$.
\vspace{2mm}

\begin{figure}[htbp]
\centering
\begin{tikzpicture}[scale=.9]
\draw[->] (0,0)--(5,0);
\draw[->] (0,0)--(0,4);
\draw[thick] plot[smooth, tension=.7] coordinates {(0.4,2.5)(1,1.5)(3,.7)(5,0.55)};
\draw[dashed,thick] plot[smooth, tension=.7] coordinates {(0.25,3.)(0.33,2.75)(0.4,2.5)};
\draw[line width=.3mm,dotted] (0.4,0)--(0.4,2.5);
\draw[line width=.3mm,dotted] (0,3)--(0.25,3);
\draw[line width=.3mm,dotted] (3,0)--(3,.7);
\node at (0.7,-0.5) {$\left(\frac{J_{\s c}}{v^{1/2}}\right)^{1/3}$};
\node at (3,-0.5) {$\frac{J_{\s c}}{v^{1/2}}$};
\draw[->] (2.3,1.8)--(2,1.25);
\draw[->] (4,1.3)--(3.9,0.75);
\node at (2.3,2.1) {Eq. $(33)$};
\node at (4.1,1.5) {Eq. $(32)$};
\draw[line width=.3mm,dotted] (0,0.55)--(5,0.55);
\node[left] at (0,3) {$\frac{1}{2}$};
\node[left] at (0,0.45) {$\sqrt{\frac{v}{\pi J_{\s c}^2}}\ln\left(\frac{J_{\s c}}{v^{1/2}}\right)$};
\node[left] at (0,3.75) {$Q_{\s LZ}$};
\node at (5,-0.5) {$v^{1/2}\tau_{\s c}$};
\end{tikzpicture}
\caption{Probability to stay on the same diabatic level after the transition is plotted schematically versus
the dimensionless correlation time, $v^{1/2}\tau_{\s c}$. In the domain of a ``true" slow noise,
$v^{1/2}\tau_{\s c} \gg J_{\s c}/v^{1/2}$, the probability,
$Q_{\s LZ}$, deviates from the asymptotic value,
$\left(\frac{v}{\pi J_{\s c}^2}\right)^{1/2}\ln\left(\frac{J_{\s c}}{v^{1/2}}\right)$, only slightly.
In the intermediate domain  $ \Big(J_{\s c}/v^{1/2}\Big)^{1/3}\ll v^{1/2}\tau_{\s c} \ll J_{\s c}/v^{1/2}$
the crossover from the slow to fast noise takes place.}
\label{fig4}
\end{figure}
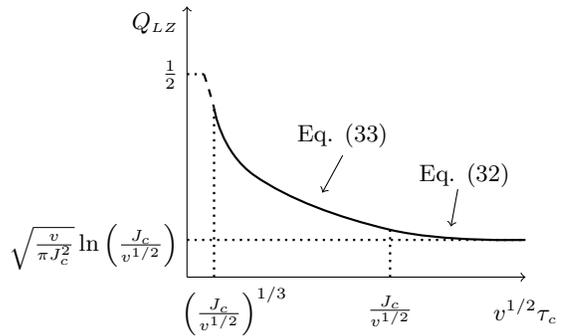

(\romannumeral 2) Large $J'$. In this limit the bracket in Eq. (\ref{Average})
should be replaced by $\left(2|J'|/v\right)^{3/2}$. Then the averaging
over $J'$ yields
\begin{equation}
\label{LargeJ}
\langle \overline{Q}_{\s LZ}\rangle \approx \frac{2}{\pi}\Gamma\big(\frac{3}{4}\big) \left(\frac{J_{\s c}}{v^{1/2}}\right)^{1/2}\left(\frac{1}{v^{1/2}\tau_{\s c}}\right)^{3/2},
\end{equation}
i.e. $\langle\overline{Q}_{\s LZ}\rangle$ grows as $\tau_{\s c}^{-3/2}$ upon the decreasing of the correlation time. The lower boundary of
the large $J'$ domain is determined by the condition $\langle\overline{Q}_{\s LZ}\rangle \sim 1$, which yields $\tau_{\s c}\sim J_{\s c}^{1/3}/v^{2/3}$.
Different behaviors of $\langle\overline{Q}_{\s LZ}\rangle$ are illustrated in Fig. \ref{fig4}.

\section{Concluding remarks}\label{Sec:Summary}

\begin{itemize}
\item[({\em \romannumeral 1})] The main results of the present paper are the expressions Eq. (\ref{Average-J}),
Eq. (\ref{Qcorrection1}), and Eq. (\ref{LargeJ}) for the probability,
$Q_{\s LZ}$, to stay on the same diabatic level after the transition.
These results pertain to the domain where this probability is small, i.e. for
$J_{\s c}^2\gg v$, or equivalently for $|\nu| \gg 1$, where the parameter
$\nu$ is defined by Eq. (\ref{nu}).  Below we combine these results into
a single expression for  $Q_{\s LZ}({\tilde \tau}_{\s c})$, where
\begin{equation}
\label{tautilde}
{\tilde \tau}_{\s c}=v^{1/2}\tau_{\s c}
\end{equation} is the dimensionless correlation time. Then one obtains
\begin{equation}
\label{final}
\langle\overline{Q}_{\s LZ}\rangle=\frac{1}{(\pi|\nu|)^{1/2}}\begin{cases}
~\ln|\nu|^{1/2}+3\cdot2^{1/2}\frac{|\nu|}{\tilde{\tau}_{\s c}^2},~\tilde{\tau}_{\s c}\gg|\nu|^{1/2}\\
\frac{2\Gamma\left(\frac{3}{4}\right) }{\pi^{1/2}} \big(\frac{|\nu|}{\tilde{\tau}_{\s c}^2}\big)^{3/4},~\!|\nu|^{1/2}\!\gg \tilde{\tau}_{\s c}\!\gg \!|\nu|^{1/6}\\
\end{cases}
\end{equation}
The second line describes the crossover between the slow-noise and  fast-noise regimes
upon decreasing the correlation time.

\begin{figure}[htbp]
\centering
\includegraphics[scale=.3]{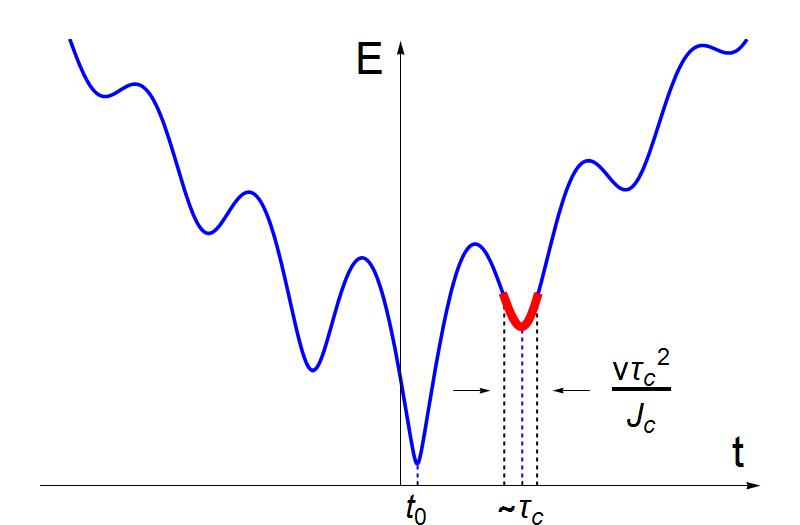}
\caption{(Color online) In the domain of correlation times $J_{\s c}^{1/3}/v^{2/3}\ll \tau_{\s c} \ll J_{\s c}/v$
a crossover between the slow-noise and fast-noise regimes takes place. In this domain, time-dependent adiabatic levels
 acquire local minima due to randomness of $J(t)$.
The duration of the Landau-Zener transition (shown with red) in the vicinity of each minimum is much shorter than $\tau_{\s c}$.
Probability to remain on the same adiabatic level after the transition, given by Eq. (\ref{multiple}), is close to $1$, so that only
the transition in the vicinity of $t=t_{\s 0}$ is responsible for $Q_{\s LZ}$.}
\label{fig5}
\end{figure}

\item[({\em \romannumeral 2})]
In obtaining the results Eqs.  (\ref{Average-J}), (\ref{Qcorrection1}),
and  (\ref{LargeJ}) we assumed that it is
only the single zero in $J(t)$ closest to the level crossing,
that is responsible for the probability $Q_{\s LZ}$. This $Q_{\s LZ}$, which is the probability for the particle to stay on the diabatic level, can also be viewed as the probability to change the adiabatic level.
In fact, in the domain $J_{\s c}^{1/3}/v^{2/3}\ll \tau_{\s c} \ll J_{\s c}/v$, many ($\sim J_{\s c}/v\tau_{\s c}$) zeros of $J(t)$ will cause local minima(maxima) in the upper(lower) adiabatic energy level, as illustrated in Fig. \ref{fig5}. In other words, many ``local" Landau-Zener transitions precede the transition near $t=0$.
For our results to apply it is necessary that, in course of each of this transition, the particle does not change the adiabatic level. Calculation of probability of changing the levels is performed in full analogy to that described above for $J(t_{\s 0})=0$ with $t_{\s 0} \ll \tau_{\s c}$. One has to linearize $J(t)$ near actual zero and perform the rotation from $\left(a_1, a_2\right)$ to  $\left(b_1, b_2\right)$. It is important that the time of the local transition turns out to be $\sim v\tau_{\s c}^2/J_{\s c}$, which is  much smaller than $\tau_{\s c}$, thus confirming that the transition is indeed local. For the probability to stay on the same adiabatic level, the above procedure yields the result
\begin{equation}
\label{multiple}
1-\exp\left[-2\pi^3\left(\frac{v^2\tau_{\s c}^3}{J_{\s c}}\right)\left(1-\frac{v^2\tau_{\s c}^2}{J_{\s c}^2}\right)\right],
\end{equation}
(actual numerical factor in the exponent depends on the location of the transition point within $\tau_{\s c}$).
For the lower boundary $\tau_{\s c}\sim J_{\s c}^{1/3}/v^{2/3}$, the number in exponent is close to $1$ indicating that at this
boundary the regime of the fast noise takes over. At the upper boundary  $\tau_{\s c}\sim J_{\s c}/v$ the number in the exponent    is large ($\sim |\nu|$). Thus the particle does not change the adiabatic level at the moments when $J(t)$ passes through zero
at times $\gtrsim \tau_{\s c}$, Fig. \ref{fig5}.

\item[({\em \romannumeral 3)}]
It is important to relate our results to the pioneering calculation
in Ref. [\onlinecite{Kayanuma1985}]. The expression for $Q_{\s LZ}$ in the presence of
 noise in this paper was based on the expansion of $Q_{\s LZ}$ in powers of $J_{\s c}^2$.

\begin{equation}
\label{Kayanuma1}
Q_{\s LZ}=\sum_{n=0}^{\infty}(-1)^nJ_{\s c}^{2n}L^{(n)},
\end{equation}
where the coefficients $L^{(n)}$ are the $2n$-fold integrals of the type
\begin{equation}
\label{Kayanuma2}
\begin{split}
L^{(n)}=& \sum\limits_{m=1}^n
\underbrace{\int_{-\infty}^{\infty}d\tau_1\int_{\tau_1}^{\infty}d\tau_2\cdots}_\text{2m}
\underbrace{\int_{-\infty}^{\infty}d\tau_{2m}\cdots\int_{-\infty}^{\tau_{2n-1}}d\tau_{2n}}_{2n-2m}\\
&\times F^{(n)}(\tau_1,\cdots,\tau_{2n})\exp\left[i\frac{v}{2}\sum_{j=1}^{2n}(-1)^j\tau_j^2\right],
\end{split}
\end{equation}
where $F^{(n)}$ is the product of the correlators $\exp\Big(|\tau_i-\tau_j|/\tau_{\s c}\Big)$ ``forbidding" the variables $\tau_i$ and $\tau_j$ to differ more than $\tau_{\s c}$.
It is seen from Eq. (\ref{Kayanuma2}) that typical values of $\tau_i$ are $v^{-1/2}$.
Thus the restrictions imposed by the correlator start to matter for the terms with $n\gtrsim \tau_{\s c} v^{1/2}$.
For smaller $n$ the restriction is not important and $F^{(n)}$ manifests itself only in the form
of combinatorial factor $  =(2n-1)!!$. With this factor in front of $L^{(n)}$,
the sum Eq. (\ref{Kayanuma1}), instead of $\exp(-2\pi J_{\s c}^2/v)$, reduces to
\begin{equation}
\label{Kayanuma3}
Q_{\s LZ}=\frac{1}{(1+4\pi J_{\s c}^2/v)^{1/2}}.
\end{equation}
The above result does not depend on the form of correlator, and is interpreted in Ref. [\onlinecite{Kayanuma1985}]
as the average of  $\exp(-2\pi J^2/v)$ with Gaussian distribution $\mathcal{P}(J)=(2\pi)^{-1/2}J_{\s c}\exp(-J^2/2J_{\s c}^2)$.
This already reveals an inconsistency, since the particular form, $\exp(-|t_i-t_j|/\tau_{\s c})$, of the correlator
chosen in Ref. [\onlinecite{Kayanuma1985}] corresponds to the telegraph noise with $\mathcal{P}(J)=\frac{1}{2}\Big(\delta(J-J_{\s c})+\delta(J+J_{\s c})\Big)$. Correspondingly, averaging over the distribution of $J$ yields $\exp(-2\pi J_{\s c}^2/v)$ instead of Eq. (\ref{Kayanuma3}), correctly implying that for $\tau_{\s c}\rightarrow \infty$ switchings of $J$ do not affect the transition probability.

For gaussian noise, our result Eq. (\ref{Average-J}) does not coincide with Eq. (\ref{Kayanuma3}). The reason is that the
logarithmic factor in Eq. (\ref{Average-J}), originating from sparse realizations, is a {\em singular} function of $J_{\s c}^2$ and cannot be captured by expansion in powers of $J_{\s c}^2$. Equally, for the telegraph noise, the finite-$\tau_{\s c}$ correction Eq. (\ref{Qcorrection}), $\sim \tau_{\s LZ}/\tau_{\s c}$, is proportional to $|J_{\s c}|$, and, thus, is also singular function of $J_{\s c}^2$. In general, the effects due to sparse realizations cannot be captured within  the
perturbative expansion.

\item[({\em \romannumeral 4})] Throughout the paper we assumed that the noise is slow and searched for
finite - $\tau_{\s c}^{-1}$ corrections
to the probability, $1-P_{\s LZ}$, to stay on the same diabatic level.
We argued that these
corrections are dominated by sparse realizations of noise.
For the fast noise the probability to stay exceeds the probability
to make a transition by $\frac{1}{2}\exp\left(-4\pi J_{\s c}^2/v\right)$.
The small parameter $\tau_{\s c}/\tau_{\s LZ}$ insures that the noise is fast.
Below we argue qualitatively
that sparse realizations of noise can
dominate the finite-$\tau_{\s c}$ correction to the standard result.

Consider the case of the fast telegraph noise. During the time $\tau_{\s LZ}$, the off-diagonal matrix element switches from $J_{\s 0}$ to $-J_{\s 0}$ approximately $\tau_{\s LZ}/\tau_{\s c}$ times. However, for certain sparse realizations, the switching happens only once at the moment $t_{\s 0}\sim \tau_{\s c}$. The probability of this realization can be estimated as $t_{\s 0}/\tau_{\s c}\exp(-\tau_{\s LZ}/{\tau_{\s c}})$. However, as was demonstrated in Sect.\ref{Sec:Telegraph}, for such realizations, the system will stay on the initial diabatic level. This suggests the following form of $P_{\s LZ}$ in the limit of fast noise
\begin{equation}
\label{eq:fast}
\frac{1}{2}-P_{\s LZ}=\frac{1}{2}\exp\left\{-\frac{4\pi J_{\s c}^2}{v}\right\}+\exp\left\{-\frac{J_{\s c}}{v\tau_{\s c}}\right\},
\end{equation}
where the second term accounts for the sparse realizations. Note that this term dominates the first term for $\tau_{\s c}\gg 1/4\pi J_{\s c}$. Remarkably, this $\tau_{\s c}$ belongs to the domain of fast noise. Indeed, for $\tau_{\s c}=1/4\pi J_{\s c}$, the ratio $\tau_{\s c}/\tau_{\s LZ}$ is equal to $v/4\pi J_{\s c}^2$, i.e. it is small. Consequently, as in the case of the slow telegraph noise, we again come to the conclusion that finite-$\tau_{\s c}$ correction can dominate the result.

\end{itemize}

\appendix

\section{Asymptotic behavior of $D_{\nu}(z)/D_{\nu}(-z)$}
\label{Sec:App}
Since we are interested in the domain $z\sim |\nu|$, standard $z\rightarrow \pm \infty$  asymptotes
of the parabolic cylinder function [\onlinecite{Bateman}] are insufficient. Therefore, we start from the following
integral representation

\begin{equation}
\label{1}
D_{\nu}(z)=\sqrt{\frac{2}{\pi}}e^{z^2/4}\int_{\s 0}^\infty e^{-t^2/2} t^\nu \cos\left(z t -\frac{\nu\pi}{2}\right)\ dt.
\end{equation}
It is convenient to divide the integral Eq. \eqref{1} into two parts
\begin{equation}
\label{2}
D_{\nu}(z)=I_+(z)e^{-i\nu\pi/2} + I_-(z)e^{i\nu\pi/2},
\end{equation}
where the functions $I_+(z)$ and $I_-(z)$ are defined as
\begin{equation}
\label{3}
I_{\pm}(z)=\sqrt{\frac{1}{2\pi}}e^{z^2/4}\int_{\s 0}^{\infty}e^{-t^2/2}t^\nu e^{\pm izt}~dt.
\end{equation}
The advantage of introducing $I_+$ and $I_-$ is that they are suited for
evaluation using the steepest descent method.
To apply this method, we rewrite $I_+$ in the form
\begin{equation}
\label{4}
I_{+}(z)=\sqrt{\frac{1}{2\pi}}e^{z^2/4}\int_{\s 0}^{\infty}e^{f(t)}\ dt,~~~~f(t)=-\frac{t^2}{2}+izt+\nu\ln t.
\end{equation}
The extrema of $f(t)$ correspond to $t_{\pm}=\frac{iz}{2}\pm \frac{\sqrt{4\nu-z^2}}{2}$. The real part of $t_+$ is positive, and thus lies within the domain of integration. By contrast, Re $t_-$ is negative and $t_-$ should therefore be excluded.

Expanding $f(t)$ near $t_+$, we get from Eq. (\ref{4})
\begin{equation}
\label{5}
\begin{split}
I_{+}(z)\approx & \sqrt{\frac{1}{2\pi}}e^{z^2/4}e^{f(t_+)} \int_{-\infty}^{\infty} d(t-t_+)\\
& \times \exp\Big[-\frac{1}{2}\Big(1+\frac{\nu}{t_+^2}\Big)(t-t_+)^2\Big].\\
\end{split}
\end{equation}
Upon performing the gaussian integration, the result can be cast in the form
 \begin{equation}
\label{6}
\begin{split}
I_{+}(z)\approx & \exp\left(\frac{iz\sqrt{4\nu-z^2}}{4}-\frac{\nu}{2}\right)(4\nu-z^2)^{-1/4}\\
&\times\ \left(\frac{iz+\sqrt{4\nu-z^2}}{2}\right)^{\nu+1/2}.\\
\end{split}
\end{equation}
The condition of applicability of the steepest descent method is that typical $(t-t_+)$ contributing to the integral Eq. (\ref{5}) is much smaller than $t_+$. It is easy to see that for $z\sim |\nu|$ this condition is satisfied when $|\nu|\gg 1$, i.e.  in the case we are interested in.
Calculation of $I_-(z)$ is similar to the above calculation. The saddle points are  $-t_{\pm}$, and only $t=-t_-$ contributes to the integral.
The result reads
\begin{equation}
\label{7}
\begin{split}
I_{-}(z)\approx &  \exp\left(-\frac{iz\sqrt{4\nu-z^2}}{4}-\frac{\nu}{2}\right)(4\nu-z^2)^{-1/4}\\
& \times \left(\frac{-iz+\sqrt{4\nu-z^2}}{2}\right)^{\nu+1/2}.\\
\end{split}
\end{equation}
Naturally, the result Eq. \eqref{7} is consistent with the relation $I_{+}(z)=I_{-}(-z)$, which follows from Eq. (\ref{3}).
If one takes a limit $z\gg |\nu|$ in Eqs.\eqref{6},\eqref{7} and substitute the result into Eq. \eqref{2}, one would recover the textbook asymptotes of $D_{\nu}(z)$.
We are interested in the ratio $D_{\nu}(-z)/D_{\nu}(z)$. Using Eqs. (\ref{5}), (\ref{6}), this ratio can be presented in a concise form
\begin{equation}
\label{8}
\begin{split}
\frac{D_\nu(-z)}{D_\nu(z)}&= \frac{I_-(z)+e^{i\nu\pi} I_+(z)}{I_+(z)+e^{i\nu\pi} I_-(z)}\\
&\approx \frac{\left(\frac{\sqrt{4\nu-z^2}-iz}{\sqrt{4\nu-z^2}+iz}\right)^{\nu+1/2}+e^{i\nu\pi+\frac{iz}{2}\sqrt{4\nu-z^2}}}{e^{\frac{iz}{2}\sqrt{4\nu-z^2}}+e^{i\nu\pi}\left(\frac{\sqrt{4\nu-z^2}-iz}{\sqrt{4\nu-z^2}+iz}\right)^{\nu+1/2}}\\
\end{split}
\end{equation}

To verify that Eq. (\ref{8}) has the right limit we recall that for $z\rightarrow \infty$ it
should reproduce $Q_{\s LZ}^{-1/2}$. Indeed, in this limit, the denominator in the fraction turns
to $\exp(\frac{iz}{2}\sqrt{4\nu-z^2})$, while the numerator turns to $\exp(\frac{iz}{2}\sqrt{4\nu-z^2}+i\nu\pi)$. Consequently, Eq. (\ref{8}) yields $\exp(\pi|\nu|)$. Furthermore, for $z\rightarrow 0$, the ratio becomes $1$, as expected.

Next we turn to the calculation of $D_{\nu-1}(-z)/D_{\nu-1}(z)$. Since the integral representation of
$D_{\nu}(z)$ contains $\nu$ in the form of $t^{\nu}$ in the integrand, evaluation of the asymptote of
$D_{\nu-1}(-z)$ with the steepest descent simply amounts to dividing the result for $I_+(z)$ by $t_+$, and the result for $I_-(z)$ by $-t_-$.
This yields
\begin{equation}
\label{9}
D_{\nu-1}(z)\approx\frac{2 e^{-i(\nu-1)\pi/2}}{iz+\sqrt{4\nu-z^2}}I_+(z)+\frac{2 e^{i(\nu-1)\pi/2}}{-iz+\sqrt{4\nu-z^2}}I_-(z).
\end{equation}
Then the ratio $D_{\nu-1}(-z)/D_{\nu-1}(z)$ can be cast in the form similar to Eq. (\ref{8})
\begin{equation}
\label{10}
\begin{split}
\frac{D_{\nu-1}(-z)}{D_{\nu-1}(z)}&= \frac{I_-(z)+e^{i\nu\pi} I_+(z)}{I_+(z)+e^{i\nu\pi} I_-(z)}\\
&\approx \frac{\left(\frac{\sqrt{4\nu-z^2}-iz}{\sqrt{4\nu-z^2}+iz}\right)^{\nu-1/2}-e^{i\nu\pi+\frac{iz}{2}\sqrt{4\nu-z^2}}}{e^{\frac{iz}{2}\sqrt{4\nu-z^2}}-e^{i\nu\pi}\left(\frac{\sqrt{4\nu-z^2}-iz}{\sqrt{4\nu-z^2}+iz}\right)^{\nu-1/2}}.\\
\end{split}
\end{equation}

%
%

%

\vspace{3mm}

\centerline{\bf Acknowledgements}

This work was supported by NSF through MRSEC (Grant No. DMR-1121252).


\begin{thebibliography}{30}
\bibitem{footnote}
    Strictly speaking, Eq. (\ref{nothing}) relates the values of $|b_1|^2$ at $t\rightarrow \pm \infty$, provided that 
    $b_2(-\infty)\rightarrow 0$. We are interested
    in the transition probability from the particular initial state $|a_1(-\infty)|^2=1$, 
    with $a_1$ being a linear combination of $b_1$ and $b_2$.
    The applicability of Eq. (\ref{nothing})
    is ensured by the fact that Eq. (\ref{phi}) determines {\em two} values of $\varphi$, which differ by $\pi/2$. 
    Transition probabilities  for these two values are the same. The proper initial state can thus be``arranged"
    by combining the solutions for $b_1$, $b_2$ corresponding to the two $\varphi$-values.
\end{thebibliography}
\end{document}